\documentclass[aps,pre,twocolumn,superscriptaddress,nofootinbib]{revtex4}

\usepackage{graphicx,amssymb,amsfonts,amsmath,color}

\newcommand{\avg}[1]{\left\langle#1\right\rangle}
\newcommand{\av}[1]{\langle#1\rangle}

\newcommand{\beq}{\begin{equation}}
\newcommand{\eeq}{\end{equation}}
\newcommand{\beqn}{\begin{eqnarray}}
\newcommand{\eeqn}{\end{eqnarray}}
\newcommand{\elabel}[1]{\label{eq:#1}}
\newcommand{\eref}[1]{Eqn.\ \ref{eq:#1}}
\newcommand{\erefs}[2]{Eqns.\ \ref{eq:#1} and \ref{eq:#2}}
\newcommand{\erefn}[2]{Eqns.\ \ref{eq:#1}-\ref{eq:#2}}
\newcommand{\flabel}[1]{\label{fig:#1}}
\newcommand{\fref}[1]{Fig.\ \ref{fig:#1}}

\newcommand{\alabel}[1]{\label{app:#1}}
\newcommand{\aref}[1]{Appendix \ref{app:#1}}

\begin{document}

\title{Limits to the precision of gradient sensing with spatial
  communication and temporal integration}

\author{Andrew Mugler}
\affiliation{Department of Physics, Purdue University}
\affiliation{Department of Physics, Emory University}

\author{Andre Levchenko}
\affiliation{Department of Biomedical Engineering and Systems Biology Institute, Yale University}

\author{Ilya Nemenman}
\email{ilya.nemenman@emory.edu}
\affiliation{Department of Physics, Emory University}
\affiliation{Department of Biology, Emory University}

\begin{abstract}
  Gradient sensing requires at least two measurements at different
  points in space. These measurements must then be communicated to a
  common location to be compared, which is unavoidably noisy. While
  much is known about the limits of measurement precision by cells,
  the limits placed by the communication are not understood. Motivated
  by recent experiments, we derive the fundamental limits to the
  precision of gradient sensing in a multicellular system, accounting
  for communication and temporal integration.  The gradient is
  estimated by comparing a ``local'' and a ``global'' molecular
  reporter of the external concentration, where the global reporter is
  exchanged between neighboring cells. Using the
  fluctuation-dissipation framework, we find, in contrast to the case
  when communication is ignored, that precision saturates with the
  number of cells independently of the measurement time duration,
  since communication establishes a maximum lengthscale over which
  sensory information can be reliably conveyed. Surprisingly, we also
  find that precision is improved if the local reporter is exchanged
  between cells as well, albeit more slowly than the global
  reporter. The reason is that while exchange of the local reporter
  weakens the comparison, it decreases the measurement noise. We term
  such a model ``regional excitation--global inhibition'' (REGI). Our
  results demonstrate that fundamental sensing limits are necessarily
  sharpened when the need to communicate information is taken into
  account.
\end{abstract}

\maketitle

Cells sense spatial gradients in environmental chemicals
with remarkable precision. A single amoeba, for example, can respond
to a difference of roughly ten attractant molecules between the front
and back of the cell \cite{Song2006}. Cells are even more sensitive
when they are in a group: cultures of many neurons respond to chemical
gradients equivalent to a difference of only one molecule across an
individual neuron's axonal growth cone \cite{Rosoff2004}; clusters of
malignant lymphocytes have a wider chemotactic sensitivity than single
cells \cite{MaletEngra:2015jc}; and groups of communicating epithelial
cells detect gradients that are too weak for a single cell to detect
\cite{us_unpublished}. More generally, collective chemosensing
properties are often very distinct from those in individual cells
\cite{Friedl:2009kz,Dona:2013cd,Pocha:2014ep,MaletEngra:2015jc}.
These observations have generated a renewed interest in the question
of what sets the fundamental limit to the precision of gradient
sensing in large, spatially extended, often collective sensory
systems.

Fundamentally, sensing a stationary gradient requires at least two
measurements to be made at different points in space. The precision of
these two or more individual measurements bounds the gradient sensing
precision \cite{Berg1977,Goodhill1999}. In their turn, each individual
measurement is limited by the finite number of molecules within the
detector volume and the ability of the detector to integrate over
time, a point first made by Berg and Purcell (BP)
\cite{Berg1977}. More detailed calculations of gradient sensing by
specific geometries of receptors have since confirmed that the
precision of gradient sensing remains limited by an expression of the
BP type \cite{Endres2009, Hu2010,Endres2008}.

However, absent in this description is the fundamental recognition
that in order for the gradient to be measured, information about
multiple spatially separated measurements must be {\em communicated}
to a common location. This point is particularly evident in the case
of multicellular sensing: if two cells at either edge of a population
measure concentrations that are different, neither cell ``knows'' this
fact until the information is shared. This is also important for a
single cell: information from receptors on either side of a cell must
be transported, e.g., via diffusive messenger molecules, to the
location of the molecular machinery that initiates the phenotypic
response. How is the precision of gradient sensing affected by this
fundamental communication requirement?

As discussed recently in the context of instantaneous measurements
\cite{us_unpublished}, the communication imposes important
limitations. First, detection of an internal diffusive messenger by
cellular machinery introduces its own BP-type limit on gradient
sensing. Since the volume of an internal detector must be smaller than
that of the whole system, and diffusion in the cytoplasm is often
slow, such an intrinsic BP limitation could be dramatic. Second, in
addition to the detection noise, the strength of the communication
itself may be hampered over long distances by messenger turnover. This
imposes a finite lengthscale over which communication is reliable,
with respect to the molecular noise. However, the communicating cells
can {\em integrate} the signals over time \cite{Berg1977}, improving
detection of even very weak messages. Whether such integration can
lift the communication constraints has not yet been addressed.

To analyze constraints on gradient sensing in spatially extended
systems with temporal integration, we use a minimal model of
collective sensing based on the local excitation--global inhibition
(LEGI) approach \cite{Levchenko2002}. This sensory mechanism uses a
local and a global internal reporter of the external concentration,
where only the global reporter is exchanged and averaged among
neighboring cells. Comparison of the two reporters then measures if
the local concentration is above or below the average, and hence
whether the cell is on the high-concentration edge of the population.
We analyze the model using a fluctuation-dissipation framework
\cite{Bialek2005}, to derive the precision with which a chemical
gradient can be estimated over long observation times.  In the case
where the need to communicate is ignored, the precision would grow
indefinitely with the number of cells. In contrast, we find that,
communication imposes limits on sensing even for long measurement
times.  Furthermore, the analysis reveals a counterintuitive strategy
for optimizing the precision.  We find that if the local reporter is
also exchanged, at a fraction of the rate of the global reporter, the
precision can be significantly enhanced. Even though such exchange
makes the two compared concentrations more similar, which weakens the
comparison, it reduces the measurement noise of the local reporter.
This tradeoff leads to an optimal ratio of exchange rates that
maximizes sensory precision. We discuss how the predictions of our
analysis could be tested experimentally.

\section{Results}
We discretize the spatially-extended gradient sensor into
compartments. These can be whole cells or their parts, but we will
refer to them as cells from now on. There is a diffusible chemical
whose concentration varies linearly in space.  The chemical gradient
defines a direction within the sensor, and we focus on a chain of
cells along this direction (\fref{cartoon}A).  Numbering the cells
from $n = 1$ to $N$, each cell experiences a local concentration
$c_n = c_N - (N-n)ag$, where $c_N$ is the background concentration,
$g$ is the concentration gradient, and $a$ is the linear size of each
cell.  We choose without loss of generality to have $g\ge 0$ and to
reference the background concentration at cell $N$, which is then at
the higher edge of the gradient.  We focus on this cell because we
imagine it will be the first to initiate a phenotypic response, such
as proliferating or directed motility. Finally, we focus on the limit
$ag/c_N\ll1$ because limits on the sensory precision will be the most
important for such small, hard to measure gradients.

\begin{figure}[tb]
	\begin{center}
		\includegraphics[width=0.5\textwidth]{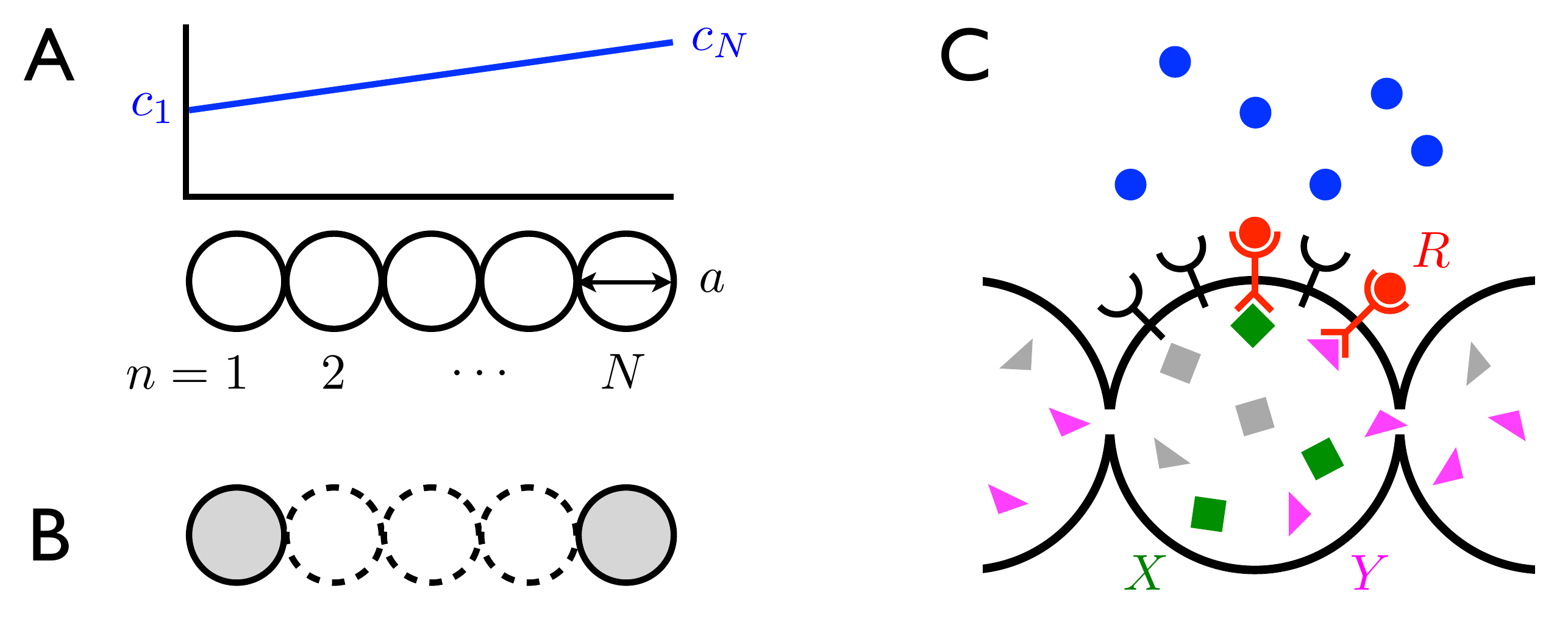}
	\end{center} 
	\caption{ \flabel{cartoon} {\bf Spatially extended gradient
            sensing.}  (A) A chain of $N$ compartments or cells is
          exposed to a linear profile of a diffusible chemical.  (B)
          In an idealized detector, the two edge cells communicate
          their measurements perfectly and instantly.  (C) In our
          model, bound receptors (R) activate a local (X) and global
          reporter molecules (Y), and Y is exchanged between cells for
          the communication.}
\end{figure}

\subsection{An idealized detector}
First, we consider the case when the two edge cells form an idealized
detector, in the sense that each cell counts every external molecule
in its vicinity, and one cell ``knows'' instantly and perfectly the
count of the other (\fref{cartoon}B).  The gradient could then be
estimated by the difference in the concentration measurements made by
the two cells \cite{Berg1977, Goodhill1999}.  The mean of this
difference is $\bar{\Delta} = \bar{c}_N-\bar{c}_1 = (N-1)ag$, and its
error is given by the corresponding errors in the two measurements,
$(\delta\Delta)^2 = (\delta c_1)^2 + (\delta c_N)^2$, under the
assumption that the measurements are independent.

Berg and Purcell showed \cite{Berg1977} that the fractional error in
each measurement is not smaller than
$(\delta c/\bar{c})^2 \sim 1/(aDT\bar{c})$, where $D$ is the diffusion
constant of the ligand, and $T$ is the time over which the measurement
is integrated.  This expression has an intuitive interpretation: the
fractional error is at least as large as the Poisson counting noise,
which scales inversely with the number of molecular counts.  The
number of counts that can be made in a time $T$ is given by the number
of molecules in the vicinity of a cell at a given time, roughly
$\bar{c}a^3$, multiplied by the number of times the diffusion renews
these molecules, $T/\tau$, where $\tau \sim a^2/D$.  This product is
$aDT\bar{c}$.

The error in the gradient estimate is then given by
$(\delta\Delta)^2 \sim \bar{c}_1/(aDT) + \bar{c}_N/(aDT)$.  For
sufficiently small gradients, such that $\bar{c}_1\approx\bar{c}_N$,
this becomes \beq \elabel{bp}
\left(\frac{\delta\Delta}{\bar{c}_N}\right)^2 \sim
\frac{1}{aDT\bar{c}_N}.  \eeq Thus, for an idealized detector, the
error in the gradient estimate is limited entirely by the error in the
measurements made by each of the edge cells. In general, we could
think of the measurement at either edge being performed by a region
that is larger than a single cell.  Since each region could not be
larger than the whole system, the highest precision is obtained when
$\sim$$Na$ replaces $a$ in \eref{bp}.  This result has been derived
more rigorously \cite{Endres2009}, and apart from a constant
prefactor, \eref{bp} indeed provides the estimation error in the limit
of large detector separation and fast detection kinetics.  More
complex geometries, such as rings of detectors \cite{Endres2009}, or
detectors distributed over the surface of a circle \cite{Hu2010} or a
sphere \cite{Endres2008}, have also been considered, and \eref{bp}
again emerges as the corresponding bound, with the lengthscale $a$
replaced by that dictated by the specific geometry.

\eref{bp} can be combined with the mean $\bar{\Delta}$ to produce the
signal-to-noise ratio (SNR) for gradient detection \beq \elabel{snr}
\frac{1}{{\rm SNR}} \equiv
\left(\frac{\delta\Delta}{\bar{\Delta}}\right)^2 \sim
\frac{\bar{c}_N}{[(N-1)ag]^2aDT} \eeq This expression again has a
clear interpretation: the SNR increases if the external molecules
diffuse more quickly ($D$) or are more sharply graded ($g$), or if the
detectors are larger ($a$), are better separated ($N$), or integrate
longer ($T$).  Yet, the SNR is worse for a larger background
concentration ($\bar{c}_N$), since it is more difficult to detect a
small gradient on top of a larger background.  The measures defined in
\erefs{bp}{snr} are conceptually equivalent only in the case of low
background concentration, when the difference $\Delta$ is comparable
to the background concentration $c_N$ (i.\ e.\ $c_1 \approx 0$).  While
much of the field has focused on \eref{bp}, here we are concerned with
the opposite case: the fundamental limits to the detection of small
gradients on a large background.  Therefore from here on we focus on
the SNR, \eref{snr}.

\subsection{Accounting for the need to communicate}
\eref{snr} cannot be a fundamental limit because it neglects a
critical aspect of gradient sensing: the need to communicate
information from multiple detectors to a common location.  Indeed, the
idealized detector implies the existence of a ``spooky action at a
distance" \cite{Mermin1985}, i.\ e.\ an unknown, instantaneous, and
error-free communication mechanism.  What are the limits to gradient
sensing when communication is properly accounted for?

To answer this, a model of gradient sensing must be assumed. A naive
model would allow each cell access to information about the input
measured and broadcast by every other cell. This would require a
number of private communication channels that grows with the number of
cells, which is not plausible. A realistic alternative that would
involve just one message being communicated is for each cell to have
access to some aggregate, average information, to which all
comparisons are made. There are a few such models
\cite{Meinhardt:1999uw,Rappel:2002fb,Jilkine:2011fy}, and our choice
among them is guided by the fact that collective detection of weak
gradients is observed in steady state, and over a wide range of
background concentration in both neurons \cite{Rosoff2004} and
epithelial cells \cite{us_unpublished}.  This supports an adaptive
spatial (rather than temporal) sensing, such as can be implemented by
the local excitation--global inhibition (LEGI) mechanism
\cite{Levchenko2002}.

The LEGI model is illustrated in \fref{cartoon}C.  Each cell contains
receptors that bind and unbind external molecules with rates $\alpha$
and $\mu$, respectively.  Bound receptors (R) activate both a local
(X) and a global (Y) intracellular species with rate $\beta$.
Deactivation of X and $Y$ occurs spontaneously with rate $\nu$.
Whereas X is confined to each cell, Y is exchanged between neighboring
cells with rate $\gamma$, which provides the cell-cell communication.
X then excites a downstream species while Y inhibits it
(LEGI). Conceptually, X measures the local concentration of external
molecules, while Y represents their spatially-averaged
concentration. If the local concentration is higher than the average
(i.e., the excitation exceeds the inhibition), then the cell is at the
higher edge of the gradient. While such comparison of the excitation
and the inhibition can be done by many different molecular mechanisms
\cite{Levchenko2002}, here we are interested in the limit of shallow
gradients. In this limit, biochemical reactions doing the comparison
can be linearized around the small difference of X and Y, and the
comparison is equivalent to subtracting Y from X
\cite{us_unpublished}.  Therefore, we take this difference, $\Delta$,
as the readout of the model.

Because we are interested in the limits to sensory precision, we focus
on the most sensitive regime, the linear response regime, where the
effects of saturation are neglected.  Introducing $r_n$, $x_n$, and
$y_n$ as the molecule numbers of R, X, and Y in the $n$th cell, the
stochastic model dynamics are
\begin{align}
\dot{c} &= D \nabla^2 c - \sum_{n=1}^N \delta(\vec{x}-\vec{x}_n)\dot{r}_n,
\nonumber\\
\dot{r}_n &= \alpha c_n - \mu r_n + \eta_n,\nonumber\\
\dot{x}_n &= \beta r_n - \nu x_n + \xi_n,\nonumber\\
\elabel{model}
\dot{y}_n &= \beta r_n - \nu \sum_{n'=1}^N M_{nn'}y_{n'} + \chi_n,
\end{align}
where $c(\vec{x},t)$ is the external concentration, and
$c_n \equiv c(\vec{x}_n,t)$ is the concentration at the location
$\vec{x}_n$ of the $n$th cell.  The matrix
$M_{nn'} \equiv (1+2\gamma/\nu)\delta_{nn'}
-(\gamma/\nu)(\delta_{n',n-1}+\delta_{n',n+1})$
includes the neighbor-to-neighbor exchange terms and is appropriately
modified at the endpoints $n \in \{1,N\}$.  $\eta_n$, $\xi_n$, and
$\chi_n$ are the noise terms. Specifically, $\eta_n$ arises from the
equilibrium binding and unbinding of external molecules to receptors
and can be expressed in terms of fluctuations in the free energy
difference $F_n$ associated with one molecule unbinding from the $n$th
cell \cite{Bialek2005}, $\eta_n = \alpha \bar{c}_n \delta F_n$ (here
$F_n$ is in units of the Boltzmann constant times temperature).
The Langevin terms $\xi_n$ and $\chi_n$ account for noise in the
activation, deactivation, and exchange reactions.  They have zero mean
and obey \cite{Gillespie:2007bx}
\begin{align}
\avg{\xi_n(t)\xi_{n'}(t')} =&\
	\delta_{n'n}(\beta \bar{r}_n + \nu \bar{x}_n)\delta(t-t'),\nonumber\\
\avg{\chi_n(t)\chi_{n'}(t')} =&\ [
	\delta_{n'n}(\beta \bar{r}_n + \nu \bar{y}_n
	+ 2\gamma \bar{y}_n + \gamma \bar{y}_{n-1} + \gamma \bar{y}_{n+1})
	\nonumber\\
\elabel{langevin}
& - \delta_{n',n-1}(\gamma \bar{y}_{n-1} + \gamma \bar{y}_n)\nonumber\\
&	- \delta_{n',n+1}(\gamma \bar{y}_{n+1} + \gamma \bar{y}_n)]\delta(t-t'),
\end{align}
where positive terms account for the Poisson noise corresponding to
each reaction, and negative terms account for the anti-correlations
introduced by the exchange.  We are particularly interested in the SNR
for the difference between local and global molecule numbers in the
edge cell, $\Delta_N = x_N - y_N$, which is the analog of \eref{snr}
for the idealized detector.

$\bar\Delta$ is given by the means $\bar{x}_N$ and $\bar{y}_N$, which
follow from \eref{model} in steady state:
$\bar{x}_N = (\beta/\nu)\bar{r}_N = [\alpha\beta/(\mu\nu)]\bar{c}_N$
and
$\bar{y}_N = (\beta/\nu)\sum_n M^{-1}_{Nn} \bar{r}_n =
[\alpha\beta/(\mu\nu)]\sum_n M^{-1}_{Nn}\bar{c}_n$,
such that \beq \elabel{signal} \bar{\Delta}_N =
\frac{\alpha\beta}{\mu\nu}\left( \bar{c}_N - \sum_{n=0}^{N-1} K_n
  \bar{c}_{N-n} \right).  \eeq Here $K_n \equiv M^{-1}_{N,N-n}$ is the
communication ``kernel'', which determines how neighboring cells'
concentration measurements are weighed in producing the global
molecule number in the edge cell.  Previously we showed
\cite{us_unpublished} that $K_n$ is \beq \elabel{K} K_n =
\frac{\sum_{j=0}^{N-n-1}{N-n-1+j \choose 2j}(\nu/\gamma)^j}
{\sum_{\ell=0}^{N-1}{N+\ell \choose 2\ell+1}(\nu/\gamma)^\ell}.  \eeq

To find the noise, we calculate the power spectra of $x_N$ and $y_N$.
As explained below, we assume that the measurement integration time
$T$ is longer than the receptor equilibration time ($\tau_1$), the
messenger turnover time ($\tau_2$), and the messenger exchange time
($\tau_3$). Under this assumption, covariances in long-time averages
are given by the low-frequency limits of the power spectra,
$C^{xy}_{nn'} = \lim_{\omega\to0} [S^{xy}_{nn'}(\omega) \equiv
\av{\tilde{\delta x}_n^* \tilde{\delta y}_{n'}}]/T$.
Linearizing \eref{model} around its means and Fourier transforming
(denoted by $\tilde{\hspace{2mm}}$) in time and space obtains
\begin{align}
-i \omega \tilde{\delta c} &= -Dk^2 \tilde{\delta c}
	+ i\omega\sum_n\tilde{\delta r}_n e^{i\vec{k}\cdot\vec{x}_n},\nonumber\\
-i\omega{\tilde\delta r}_n &= \alpha\hat{\delta c}(\vec{x}_n,\omega)
	- \mu\tilde{\delta r}_n + \alpha\bar{c}_n \tilde{\delta F}_n,\nonumber\\
-i\omega\tilde{\delta x}_n &= \beta \tilde{\delta r}_n
	- \nu \tilde{\delta x}_n + \tilde{\xi}_n,\nonumber\\
\elabel{fourier}
-i\omega\tilde{\delta y}_n &= \beta \tilde{\delta r}_n
	- \nu \sum_{n'} M_{nn'}\tilde{\delta y}_{n'}
	+ \tilde{\chi}_n,
\end{align}
where
$\hat{\delta c}(\vec{x},\omega) \equiv \int
d^3k\,(2\pi)^{-3}\tilde{\delta
  c}(\vec{k},\omega)e^{-i\vec{k}\cdot\vec{x}}$.
The first step in finding the noise is to calculate the power spectrum for $r_n$, which we
do using the fluctuation-dissipation theorem (FDT) as in
\cite{Bialek2005}.  FDT relates the power spectrum
$S^{rr}_{nn'}(\omega)$ (fluctuations) to the imaginary part of the
generalized susceptibility $G_{nn'}(\omega)$ (dissipation), \beq
\elabel{fdt} S^{rr}_{nn'}(\omega) = \frac{2}{\omega}{\rm
  Im}[G_{nn'}(\omega)], \eeq where $G_{nn'}(\omega)$ describes how the
receptor binding relaxes to small changes in the free energy, \beq
\elabel{susceptibility} \tilde{\delta r}_n = \sum_{n'} G_{nn'}(\omega)
\tilde{\delta F}_{n'}.  \eeq We solve for $G_{nn'}(\omega)$ by
eliminating $\tilde{\delta c}$ from the first two lines of
\eref{fourier}, which yields a relationship between
$\tilde{\delta r}_n$ and $\tilde{\delta F}_n$.  As detailed in \aref{srr},
writing this relationship in the form of
\eref{susceptibility} requires inverting a Toeplitz marix (a matrix
with constant diagonals), which has a known inversion algorithm
\cite{Zohar1969}.
The result is \beq \elabel{Sr} S^{rr}_{nn'}(\omega) =
\frac{2\alpha\bar{c}_{n'}}{\mu^2}
\begin{cases}
\left(1+\frac{\alpha}{2\pi aD}\right)
	& n'=n,\\
\frac{\alpha}{4\pi aD}\frac{1}{|n-n'|}
	& n'\ne n.
\end{cases}
\eeq Here the cell diameter $a$ appears because we cut off the
wavevector integrals at the maximal value $k \sim \pi/a$, as in
\cite{Bialek2005}. This regularizes unphysical divergences caused by
the $\delta$-correlated noises in the Langevin approximation in
\eref{model}. In deriving \eref{Sr}, we have made the first of our
timescale assumptions, namely
$T \gg \tau_1 \equiv \mu^{-1} + K/4\pi\sigma D$, where
$K \equiv \alpha/\mu$ is the equilibrium constant, and
$\sigma \equiv a/2$ is the cell radius. $\tau_1$ is the receptor
equilibration timescale: it is the time it takes for a signal molecule
to unbind from the receptors and diffuse away from the cell into the
bulk \cite{Agmon1990}. Its first term is the intrinsic receptor
unbinding time, and its second term accounts for rebinding events
before the molecule diffuses far away \cite{Kaizu2014}.

The second step is to calculate power spectra for $x_N$ and $y_N$
using the last two lines of \eref{fourier},
\begin{align}
S_{NN}^{xx}(\omega) &= \frac{1}{\nu^2+\omega^2}
	\left[\beta^2 S_{NN}^r(\omega) + \avg{\tilde{\xi}_N^*\tilde{\xi}_N}\right], \nonumber\\
S_{NN}^{yy}(\omega) &= \frac{1}{\nu^2}\sum_{nn'}
	\tilde{M}_{Nn}^{-1*}\tilde{M}_{Nn'}^{-1}
	\left[\beta^2 S_{nn'}^r(\omega) + \avg{\tilde{\chi}_{n}^*\tilde{\chi}_{n'}}\right], \nonumber \\
\elabel{Sxy}
S_{NN}^{xy}(\omega) &= \frac{1}{\nu(\nu+i\omega)}\sum_n\tilde{M}^{-1}_{Nn}\beta^2S^r_{Nn}(\omega),
\end{align}
where $\tilde{M}_{nn'} \equiv M_{nn'}-i(\omega/\nu)\delta_{nn'}$.  Now
taking the low-frequency limit imposes our second timescale
assumption, namely $T \gg \tau_2 \equiv 1/\nu$, where $\tau_2$ is the
timescale of messenger turnover by degradation.  The noise spectra in
\eref{Sxy} follow directly from Fourier transforming \eref{langevin}
and using the steady state means of \eref{model} to eliminate
$\beta\bar{r}_n$,
\begin{align}
\avg{\tilde{\xi}_N^*\tilde{\xi}_N} &= 2 \nu \bar{x}_N, \nonumber \\
\elabel{xichi}
\avg{\tilde{\chi}_{n}^*\tilde{\chi}_{n'}} &= \nu (M_{nn'}\bar{y}_{n'} + M_{n'n}\bar{y}_n).
\end{align}
The appearance of $M_{nn'}$ in \eref{xichi} is expected, since the
noise arises in reactions in every cell, and then propagates to other
cells via the same matrix as the means \cite{Detwiler2000}.  Indeed,
this simplifies the second line in \eref{Sxy} for $\omega\to0$, since
then $M_{nn'} = \tilde{M}_{nn'}(\omega = 0)$ hits its own inverse.
The result is an expression for the variance
$(\delta\Delta_N)^2 = (\delta x_N)^2 + (\delta y_N)^2 - 2C^{xy}_{NN} =
[S_{NN}^{xx}(0) + S_{NN}^{yy}(0) - 2S_{NN}^{xy}(0)]/T$, namely
\begin{multline}
(\delta\Delta_N)^2 =\frac{\beta^2}{\nu^2}\left[\frac{S^{rr}_{NN}(0)}{T}
	+ \sum_{nn'} K_{N-n} K_{N-n'} \frac{S^{rr}_{nn'}(0)}{T} \right. \\
\elabel{noise}
	\left. -\ 2\sum_n K_{N-n} \frac{S^{rr}_{Nn}(0)}{T} \right]
	+ \frac{2}{\nu T}\left( \bar{x}_N + K_0\bar{y}_N \right).
\end{multline}
\erefs{signal}{noise}, together with \erefs{K}{Sr}, give the
${\rm SNR} = (\bar{\Delta}_N/\delta\Delta_N)^2$, which we do not write
here for brevity.

The SNR is compared with the result for the idealized detector
(\eref{snr}) in \fref{snr}.  We see that whereas the SNR for the
idealized detector increases indefinitely with the number of cells
$N$, the SNR for the model with communication {\em and} temporal
integration saturates, as in the no-integration case
\cite{us_unpublished}.  This is our first main finding: communication
leads to a maximum precision of gradient sensing, which a
multicellular system cannot surpass no matter how large it grows or
how long it integrates.  The reason is that communication is not
infinitely precise over large lengthscales.  In the next section, we
make this point clear by deriving a simple fundamental expression for
the maximum value of the SNR.

\begin{figure}[tb]
	\begin{center}
		\includegraphics[width=0.47\textwidth]{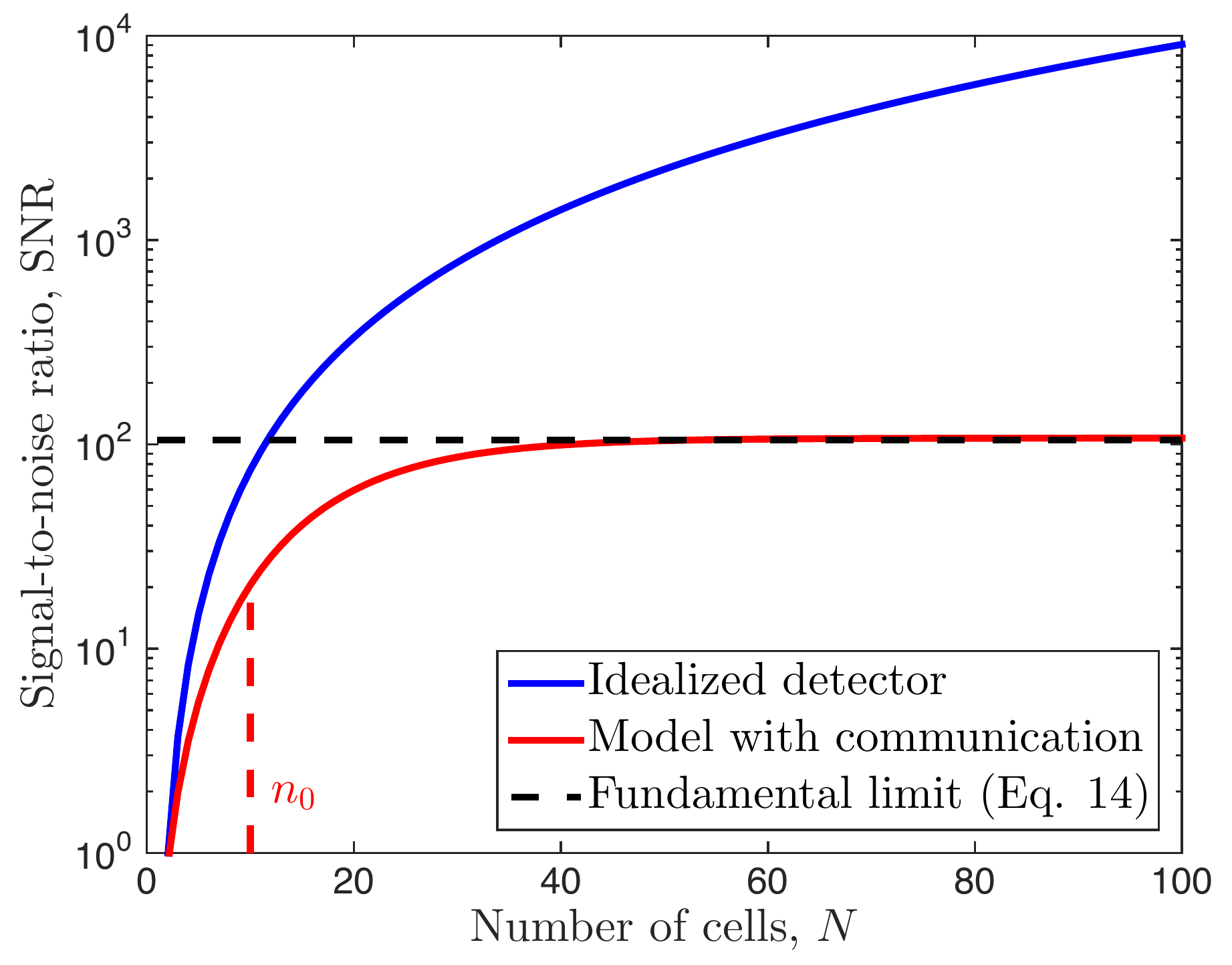}
	\end{center}
	\caption{ \flabel{snr} {\bf Precision of gradient sensing with
            temporal integration}. Signal-to-noise ratio (SNR) vs.\
          number of cells $N$ is shown for the idealized detector
          (\eref{snr} with prefactor $1/\pi$) and for our model with
          communication (Eqs.\ \ref{eq:signal}, \ref{eq:K},
          \ref{eq:Sr}, and \ref{eq:noise}). Whereas the SNR for the
          idealized detector increases indefinitely, the SNR for the
          model with communication saturates for $N \gg n_0$. The
          saturation level is bounded from above by the fundamental
          limit, \eref{limit}. As shown, the bound is reached in the
          high-gain regime $\alpha/a^3\mu = \beta/\nu = 100$, where
          intrinsic noise is negligible. Other parameters are
          $a = 10\ \mu$m, $\bar{c}_N = 1$ nM, $g = 1$ nM/mm,
          $D = 50\ \mu$m$^2$/s, $\mu = \nu = 1$ s$^{-1}$,
          $n_0 = \sqrt{\gamma/\nu} = 10$, and the integration time
          scale is $T = 10$ s.  }
\end{figure}

\subsection{Fundamental limit to sensory precision}

The saturating value of the SNR is obtained in the limit of large $N$.
In this limit, and when communication is strong ($\gamma \gg \nu$),
the kernel (\eref{K}) reduces to $K(n,n_0) \approx e^{-n/n_0}/n_0$
\cite{us_unpublished}.  Here $n_0 \equiv \sqrt{\gamma/\nu}$ sets the
lengthscale of the kernel and therefore sets the number of neighboring
cells with which the edge cell effectively communicates.
The limit $\gamma \gg \nu$ and our assumption $T \gg \tau_2 = 1/\nu$ imply our third timescale assumption, $T \gg \tau_3 \equiv 1/\gamma$, i.e.\ that the integration time is longer than the timescale of messenger exchange from cell to cell.
Inserting
the expression for $K(n,n_0)$ into \eref{signal}, and approximating the
sum as an integral in the large $N$ limit, the mean becomes
$\bar{\Delta}_N \approx
(\alpha\beta/\mu\nu)(\bar{c}_N-\bar{c}_{N-n_0}) = \alpha\beta
n_0ag/\mu\nu$.
Inserting $K(n,n_0)$ into \eref{noise} results in products of the
exponential with the $1/|n-n'|$ dependence of the bound receptor power
spectrum (\eref{Sr}), leading to sums of the type $\sum_je^{-j}/j$,
which we evaluate in \aref{limit}.  The result is 
\begin{align} \elabel{limit}
\frac{1}{{\rm SNR}} &=
\left(\frac{\delta\Delta_N}{\bar{\Delta}_N}\right)^2 \gtrsim
\frac{c_{\rm eff}}{\pi (n_0ag)^2aDT}, \;{\rm where}\\
\elabel{ceff} c_{\rm eff} &\equiv \bar{c}_N + \frac{\log n_0}{2n_0}
(\bar{c}_{N-n_0/2} - 2\bar{c}_N),  
\end{align}
and $T \gg \{\tau_1,\tau_2,\tau_3\}$.  \eref{limit} is fundamental in
the sense that it does not depend on the details of the internal
sensory mechanism. Rather, it only depends on the properties of the
external signal ($c,g,D$), the physical dimensions ($a$), and the fact
that information is integrated ($T$) and communicated ($n_0$) by the
cells.
The inequality reflects the fact that the righthand side contains
additional positive terms arising from the finite number of bound
receptors and intracellular molecules (see \aref{limit}).  These terms represent
intrinsic noise and can in principle be made arbitrarily small by
increasing the gain factors $\alpha/a^3\mu$ and $\beta/\nu$, which
dictate the internal molecule numbers.  What remains in \eref{limit}
is the communicated extrinsic noise, which arises unavoidably from the
diffusive fluctuations in the numbers of the ligand molecules being
detected. \eref{limit} is shown to bound the exact SNR in \fref{snr}.

Comparing \eref{limit} to the expression for the idealized detector
(\eref{snr}), we see that the expressions are very similar but contain
two important differences.  First, whereas \eref{snr} decreases
indefinitely with $N$, \eref{limit} remains bounded by $n_0$ for large
$N$ (see \fref{snr}).  Evidently, a very large detector is limited in its precision to
that of a smaller detector with effective size $n_0$.  This limitation
reflects the fact that reliable communication is restricted to a
finite lengthscale.  Importantly, \eref{limit} demonstrates that this
noise is present independently of the number of intrinsic signaling
molecules in the communication channel. Thus the fundamental sensory
limit is affected not only by the measurement process (as in BP
theory), but also unavoidably by the communication process.

The second important difference is that \eref{snr} depends on
$\bar{c}_N$, whereas \eref{limit} depends on the effective
concentration $c_{\rm eff}$, defined in \eref{ceff}. $c_{\rm eff}$ is
a sum of the concentration measured by the local species, by the
global species, and the covariance between them, respectively. The
local species measures the concentration only within its local
vicinity, $\bar{c}_N$. However, the global species effectively
measures the concentration in the vicinity of $n_0$ cells. This fact
reduces the noise associated with this term (and the covariance term)
by the factor of $n_0$ in the denominator of \eref{ceff}. Because
intercellular molecular exchange also competes with extracellular
molecular diffusion, not all of the measurements made by these $n_0$
cells are independent. Therefore, the reduction is tempered by the
$\log n_0$ factor in the numerator of \eref{ceff}. This log arises
from the interaction of the $e^{-n}$ exchange kernel with the
$1/|n-n'|$ diffusion kernel (\aref{limit}). The net result is that, because
of the correlations imposed by externals diffusion, the number of
independent measurements grows sublinearly with the system
size. Nonetheless, it does grow, and, correspondingly, in \eref{ceff}
the measurement noise in the global species decreases with the
communication lengthscale $n_0$.  Crucially, this means that
\eref{limit} is dominated by the measurement noise of the local
species, i.e.\ the first term in \eref{ceff}.

In deriving the precision of gradient sensing, we have also derived
the precision of concentration sensing by communicating
cells. Specifically, by focusing only on the global species terms, and
following the steps leading to \eref{limit}, we get
\begin{equation}
\elabel{conc}
  \left(\frac{\delta y_N}{\bar{y}_N}\right)^2\gtrsim \frac{1}{2\pi a_{\rm eff}DTc_{\rm eff}}, 
\end{equation}
where $a_{\rm eff} \equiv an_0/\log n_0$ and
$c_{\rm eff} \equiv \bar{c}_{N-n_0}^2/\bar{c}_{N-n_0/2}$. This
expression has the same form as the BP limit,
$(\delta c/\bar{c})^2 \sim 1/(aDT\bar{c})$. Indeed, in the absence of
a gradient, $\bar{c}_n = \bar{c}_N$ is constant, and
$c_{\rm eff} \to \bar{c}_N$. Importantly, however, the effect of
communication remains present in $a_{\rm eff}$: messenger exchange
expands the effective detection lengthscale by a factor $n_0$, while
ligand diffusion once again tempers the expansion by $\log n_0$. The
net result is that communication reduces error by increasing the
effective detector size, $a_{\rm eff} > a$.

\subsection{Optimal sensing strategy}

In the previous section, we saw that the limit to the precision of
multicellular gradient sensing is dominated by the measurement noise
of the local species in the edge cell (\erefs{limit}{ceff}). In
contrast, the measurement noise of the global species is reduced by
the intercellular communication. This finding raises an interesting
question: could the total noise be further reduced if the local
species were also exchanged between cells? To explore this
possibility, we extend the model in \erefs{model}{langevin} to allow
for exchange of the local species at rate $\gamma_x$, and we take
$\gamma \to \gamma_y > \gamma_x$ for the global species. An immediate
consequence of this modification is that the signal becomes
$\bar{\Delta}_N \approx \alpha\beta (n_y-n_x) ag/\mu\nu$, where
$n_x \equiv \gamma_x/\nu$ and $n_y = \gamma_y/\nu$. Thus the signal is
reduced by local exchange, since increasing $\gamma_x$ decreases the
difference $n_y-n_x$. This is because the signal is defined by the
difference between the global and local readouts, and allowing for the
local species exchange makes the two readouts less different. We thus
anticipate that any useful local exchange rate will satisfy
$\gamma_x \ll \gamma_y$ to maintain sufficiently high signal. In this
limit, we find that \eref{ceff} remains dominated by the first term,
even as local exchange reduces this term according to
$c_{N} \to \bar{c}_N(\log n_x)/2n_x$. \eref{limit} then becomes \beq
\elabel{snr2} {\rm SNR} \lesssim
\frac{2\pi(ag)^2aDT}{\bar{c}_N}\frac{(n_y-n_x)^2n_x}{\log n_x}.  \eeq
For large $n_y$, this expression has a maximum as a function of
$n_x$. The maximum arises due to a fundamental tradeoff: exchange of
the local species reduces the signal, but it also reduces the dominant
local measurement noise.

The optimal value $n_x^*$ depends on $n_y$. Experiments in epithelial
cells suggest that the communication lengthscale $n_y$ is on the order
of a few to ten cells \cite{us_unpublished}. In this range, we find
the optimum numerically from the {\em exact} SNR, which comes from
straightforwardly generalizing \eref{noise} (\aref{regi}).
\fref{regi} shows that $n_x^*$ is about half of $n_y$ for one
specific set of parameters, leading to an optimal exchange rate ratio
of $\gamma_x^*/\gamma_y = (n_x^*/n_y)^2 \approx 25\%$. While the exact
optimal ratio depends on the relative strengths of different noises,
and hence on the gains (see \fref{gain} in \aref{regi}), the main finding is robust: a
multicellular system should exchange both antagonistic messenger
molecules, one at a fraction of the rate of the other.  We call this
strategy {\em regional} excitation--global inhibition (REGI).
\fref{regi} shows that the enhancement over the one-messenger LEGI
strategy can be substantial.  For example, with $n_y = 10$ cells, the
SNR is optimally enhanced by a factor of $5$. With $n_y = 15$, the
enhancement is almost $8$-fold.

\begin{figure}[tb]
	\begin{center}
          \includegraphics[width=0.47\textwidth]{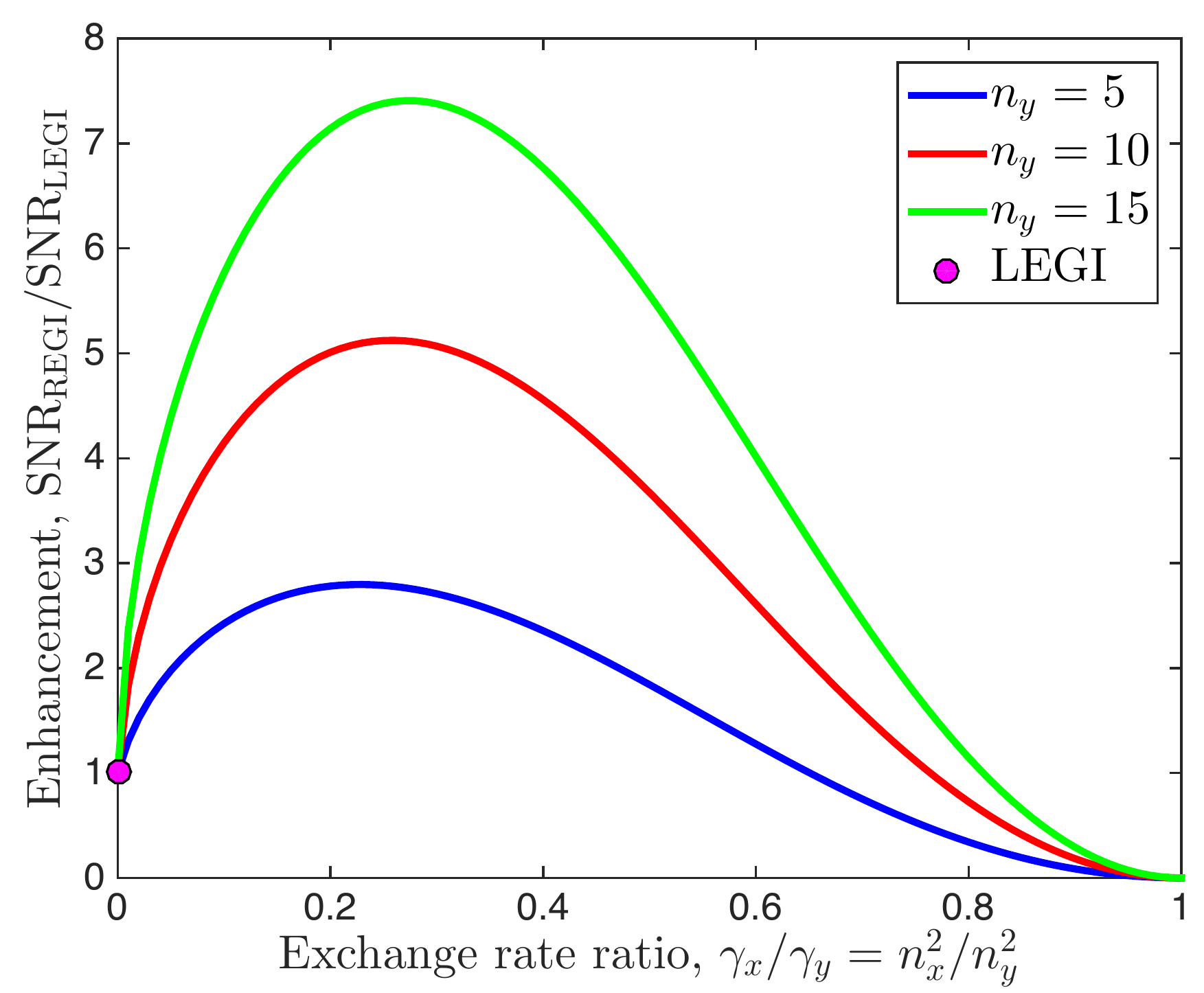}
	\end{center}
	\caption{ \flabel{regi}{\bf Regional excitation--global
            inhibition (REGI).} Signal-to-noise ratio (SNR) is
          enhanced by allowing both messengers to be exchanged between
          cells. The optimal enhancement over LEGI is substantial and
          occurs because exchange of the local species reduces
          measurement noise, but also reduces the signal. Parameters
          are as in \fref{snr}, but with $N=100$,
          $\alpha/a^3\mu = \beta/\nu = 5$ and several values of
          $n_y \equiv \sqrt{\gamma_y/\nu}$ as indicated.  }
\end{figure}

\section{Discussion}

Cellular sensing of spatially inhomogeneous concentrations is a
fundamental biological computation, involved in a variety of processes
in the development and behavior of living systems. Like binocular
vision and stereophonic sound processing, it is a process where the
sensing is done by an array of spatially distributed sensors. Thus the
accuracy of sensing is limited in part by the physical properties of
the biological machinery that brings together the many spatially
distributed measurements. Understanding these limits is a difficult
problem.

Here we solved this problem in the case where the communication is
diffusive, and one-dimensional distributed measurements are used to
calculate external concentration gradients within the LEGI
paradigm. We allowed for temporal integration, extending the results
of Ref.~\cite{us_unpublished}. Some of the features of the gradient
sensing limit we derived (\eref{limit}), such as the unbounded
increase of the SNR with the diffusion coefficient of the ligand or
with the integration time, carry over from the Berg-Purcell theory of
gradient sensing \cite{Endres2008}, which does not account for
communication. However, our most important finding is that, in
contrast to the BP theory, the growth of the sensor array beyond a
certain size stops increasing the SNR. The effect is independent of
the intrinsic noise in the communication system and thus represents a
truly fundamental limitation of diffusive communication for
distributed sensing. In particular, it holds for multicellular
systems, as well as for large individual cells.  Although we derived
the limit for a linear signal profile, we anticipate that the limit
for a nonlinear profile will be similar, \eref{limit}, but with a
different effective concentration $c_{\rm eff}$.  It remains to be
seen if similar limits hold when the sensors are arranged in two- or
three-dimensional structures, or when concentration information
propagates super-diffusively, as is possible in wave-based or
Turing-type models of polarization establishment
\cite{Jilkine:2011fy}. Additionally, it will be important to relax
various assumptions of the model, such as allowing for saturation of
receptors or limiting the total number of messengers, and coupling the
model to the motility apparatus to investigate how the improved
sensory precision affects downstream functions.

Our results illustrate two important features of temporal averaging by
distributed sensors. First, our derivation naturally reveals which
timescales are relevant in this process, namely receptor equilibration
($\tau_1 = \mu^{-1} + K/4\pi\sigma D$), messenger turnover
($\tau_2 = 1/\nu$), and messenger exchange ($\tau_3 = 1/\gamma$). In
principle, these timescales could have depended on system-level
properties, such as the system size ($N$) or the communication length
($n_0$). Surprisingly, instead they depend only on single-cell
properties, meaning that efficient temporal averaging is not slowed
down by increasing the number of sensors. Second, our results reveal
the effects of over-counting due to correlations between external and
internal diffusion. In both gradient sensing (\erefs{limit}{ceff}) and
concentration sensing (\eref{conc}), we see that the noise reduction
afforded by communication-based averaging is tempered by a factor
$\log n_0$. This log is not a mathematical curiosity. Rather, it
reflects the fact that not all measurements communicated to a cell by
its neighbors are independent since the signal molecules also diffuse
externally. Coupling external diffusion with internal exchange
introduces correlations among measurements, which reduces the benefit
of internal averaging.

Another central prediction is that the gradient sensing is improved by
a system with two messengers, exchanged at different rates. We call
this mechanism regional excitation--global inhibition (REGI), a
generalization of the standard local excitation--global inhibition
(LEGI) model. Optimality of REGI follows directly from the interplay
between the ligand stochasticity and the communication
constraints. Therefore, REGI has not been identified as an optimal
strategy in previous studies that neglected either of these two
effects. However, evidence for REGI may already exist in large
gradient-sensing cells, where activated receptor complexes, which
diffuse in the membrane at a rate $\sim$$10$$-$$100$ times slower than
similar cytosolic molecules \cite{kholodenko}, may act as the regional
messengers.

REGI emerged from maximizing the SNR in our system, which revealed the
optimal rate ratio $\gamma_x/\gamma_y$. Maximization of the SNR also
implies that the optimal value of $\gamma_y$ (or $n_y$) is infinity,
since the SNR grows indefinitely with $n_y$ (see
\fref{regi}). Infinite $n_y$ corresponds to averaging over as large a
distance as possible. Such a strategy is only optimal here because the
concentration profile is linear, with constant gradient $g$. In
contrast, more physical nonlinear profiles (e.g.\ exponential,
power-law, randomly varying, or profiles with extrema) have spatially
varying $g$. In these cases, if the size of the group of cells is
larger than the correlation length of $g$, then an infinite $n_y$
would average out the signal together with the noise, which would
reduce the SNR. In contrast, a finite $n_y$ would allow a subset of
cells to detect the local gradient in their vicinity, which is an
essential task in morphological processes such as tissue branching and
collective migration.

REGI can be interpreted as performing a spatial
derivative. Specifically, the two-lobed filter $K(n,n_x)-K(n,n_y)$
reports the difference in concentrations measured over distances $n_x$
and $n_y$ near a given detector, and the values $n_x$ and $n_y$ depend
on the properties of the environment. Thus REGI is similar to the
temporal differentiation in {\em E.~coli} chemotaxis. Indeed, the
temporal filter of the {\em E.~coli} sensory module is also two-lobed,
with the short and long timescales set by ligand statistics and
rotational diffusion, respectively \cite{Berg:2004wy}.  Thus, for both
spatial and temporal filtering, the choice of the two optimal length-
or timescales is determined by matching the filter to the statistical
properties of the signal and the noise \cite{Wiener}, which is
understood well for {\em E.~coli} \cite{Berg:2004wy}.

Interpreting the REGI model as a spatiotemporal filter
suggests experiments that would identify if a certain biological
system employs this mechanism. Such experiments would involve
concentration profiles that differ substantially from steady-state
linear gradients.
For example, subjecting cells to a concentration profile with a
spatially localized maximum would allow one to measure both $\gamma_x$
and $\gamma_y$ by observing the response of cells near the
concentration peak as a function of the peak width. Alternatively, one
can subject cells to a spatiotemporally localized concentration pulse
and observe if a response a certain distance away from the pulse
exhibits the signs of only inhibition (LEGI, one messenger), or
inhibition and excitation on different scales (REGI, two messengers).
Understanding fundamental sensory limits for diffusive communication
in gradient sensing opens up possibilities to propose and analyze
these and other related experiments.

\begin{acknowledgments}
  We thank Matt Brennan for useful discussions. AM and IN were
  supported in part by the James S.\ McDonnell Foundation grant
  220020321, and the Human Frontiers Science Program grant
  RGY0084/2011. IN was further supported by the NSF grant
  PoLS-1410978. AL was supported in part by NIH grants CA155758 and
  GM072024, by NSF grant PoLS-1410545, and by the Semiconductor
  Research Corporation's SemiSynBio program.
\end{acknowledgments}

\bibliographystyle{pnas}
\bibliography{refs}

\begin{thebibliography}{10}

\bibitem{Song2006}
Song L, {et~al.}
\newblock (2006) Dictyostelium discoideum chemotaxis: threshold for directed
  motion.
\newblock \emph{Eur J Cell Biol} 85:981--9.

\bibitem{Rosoff2004}
Rosoff WJ, {et~al.}
\newblock (2004) A new chemotaxis assay shows the extreme sensitivity of axons
  to molecular gradients.
\newblock \emph{Nat Neurosci} 7:678--82.

\bibitem{MaletEngra:2015jc}
Malet-Engra G, {et~al.}
\newblock (2015) {Collective cell motility promotes chemotactic prowess and
  resistance to chemorepulsion.}
\newblock \emph{Curr Biol} 25:242--250.

\bibitem{us_unpublished}
Ellison D, {et~al.}
\newblock (2015) Cell-cell communication can enhance the effect of shallow
  gradients of cues guiding cell growth and morphogenesis.
\newblock Submitted.

\bibitem{Friedl:2009kz}
Friedl P, Gilmour D
\newblock (2009) {Collective cell migration in morphogenesis, regeneration and
  cancer.}
\newblock \emph{Nat Rev Mol Cell Biol} 10:445--457.

\bibitem{Dona:2013cd}
Don{\`a} E, {et~al.}
\newblock (2013) {Directional tissue migration through a self-generated
  chemokine gradient.}
\newblock \emph{Nature} 503:285--289.

\bibitem{Pocha:2014ep}
Pocha SM, Montell DJ
\newblock (2014) {Cellular and molecular mechanisms of single and collective
  cell migrations in Drosophila: themes and variations.}
\newblock \emph{Annual review of genetics} 48:295--318.

\bibitem{Berg1977}
Berg HC, Purcell EM
\newblock (1977) Physics of chemoreception.
\newblock \emph{Biophys J} 20:193--219.

\bibitem{Goodhill1999}
Goodhill GJ, Urbach JS
\newblock (1999) Theoretical analysis of gradient detection by growth cones.
\newblock \emph{J Neurobiol} 41:230--41.

\bibitem{Endres2009}
Endres RG, Wingreen NS
\newblock (2009) Accuracy of direct gradient sensing by cell-surface receptors.
\newblock \emph{Progr Biophys Mol Biol} 100:33--9.

\bibitem{Hu2010}
Hu B, Chen W, Rappel WJ, Levine H
\newblock (2010) Physical limits on cellular sensing of spatial gradients.
\newblock \emph{Physical Review Letters} 105:48104.

\bibitem{Endres2008}
Endres RG, Wingreen NS
\newblock (2008) Accuracy of direct gradient sensing by single cells.
\newblock \emph{Proc Natl Acad Sci USA} 105:15749--54.

\bibitem{Levchenko2002}
Levchenko A, Iglesias PA
\newblock (2002) Models of eukaryotic gradient sensing: application to
  chemotaxis of amoebae and neutrophils.
\newblock \emph{Biophysical Journal} 82:50--63.

\bibitem{Bialek2005}
Bialek W, Setayeshgar S
\newblock (2005) Physical limits to biochemical signaling.
\newblock \emph{Proc Natl Acad Sci USA} 102:10040--5.

\bibitem{Mermin1985}
Mermin N
\newblock (1985) Is the moon there when nobody looks? reality and the quantum
  theory.
\newblock \emph{Physics Today} 38:38--47.

\bibitem{Meinhardt:1999uw}
Meinhardt H
\newblock (1999) {Orientation of chemotactic cells and growth cones: models and
  mechanisms.}
\newblock \emph{J Cell Sci} 112 ( Pt 17):2867--2874.

\bibitem{Rappel:2002fb}
Rappel WJ, Thomas PJ, Levine H, Loomis WF
\newblock (2002) {Establishing direction during chemotaxis in eukaryotic
  cells}.
\newblock \emph{Biophys J} 83:1361--1367.

\bibitem{Jilkine:2011fy}
Jilkine A, Edelstein-Keshet L
\newblock (2011) {A comparison of mathematical models for polarization of
  single eukaryotic cells in response to guided cues.}
\newblock \emph{PLoS Comp Biol} 7:e1001121.

\bibitem{Gillespie:2007bx}
Gillespie DT
\newblock (2007) {Stochastic simulation of chemical kinetics.}
\newblock \emph{Ann Rev Phys Chem} 58:35--55.

\bibitem{Zohar1969}
Zohar S
\newblock (1969) Toeplitz matrix inversion: the algorithm of {WF} {T}rench.
\newblock \emph{J ACM} 16:592--601.

\bibitem{Agmon1990}
Agmon N, Szabo A
\newblock (1990) Theory of reversible diffusion-influenced reactions.
\newblock \emph{J Chem Phys} 92:5270--5284.

\bibitem{Kaizu2014}
Kaizu K, {et~al.}
\newblock (2014) The {B}erg-{P}urcell limit revisited.
\newblock \emph{Biophys J} 106:976--985.

\bibitem{Detwiler2000}
Detwiler PB, Ramanathan S, Sengupta A, Shraiman BI
\newblock (2000) Engineering aspects of enzymatic signal transduction:
  photoreceptors in the retina.
\newblock \emph{Biophysical Journal} 79:2801--2817.

\bibitem{kholodenko}
Kholodenko B, Hoek J, Westerhoff H
\newblock (2000) Why cytoplasmic signalling proteins should be recruited to
  cell membranes.
\newblock \emph{Trends Cell Biol} 10:173.

\bibitem{Berg:2004wy}
Berg HC
\newblock (2004) \emph{{E. Coli in Motion}}
\newblock (Springer).

\bibitem{Wiener}
Wiener N
\newblock (1964) \emph{Extrapolation, Interpolation, and Smoothing of
  Stationary Time Series: With Engineering Applications}
\newblock (MIT Press).

\end{thebibliography}

\onecolumngrid
\appendix

\section{Power spectrum of the bound receptor number}
\alabel{srr}
The dynamics of receptor binding are given in Fourier space by the
first two lines of \eref{fourier},
\begin{align}
\elabel{c}
-i \omega \tilde{\delta c} &= -Dk^2 \tilde{\delta c}
	+ i\omega\sum_n\tilde{\delta r}_n e^{i\vec{k}\cdot\vec{x}_n}, \\
\elabel{r}
-i\omega\tilde{\delta r}_n &= \alpha\hat{\delta c}(\vec{x}_n,\omega)
	- \mu\tilde{\delta r}_n + \alpha\bar{c}_n \tilde{\delta F}_n,
\end{align}
where
\beq
\elabel{chat}
\hat{\delta c}(\vec{x},\omega) \equiv \int d^3k\,(2\pi)^{-3}\tilde{\delta c}(\vec{k},\omega)e^{-i\vec{k}\cdot\vec{x}}.
\eeq
We solve \eref{c} for $\tilde{\delta c}$ and, using \eref{chat}, insert it into \eref{r} to obtain
\beq
\elabel{rf}
\{\mu-i\omega[1+\alpha\Sigma(\omega)]\}\tilde{\delta r}_n
- i\omega\alpha\sum_{n'\neq n}V(|\vec{x}_n-\vec{x}_{n'}|,\omega)\tilde{\delta r}_{n'}
= \alpha\bar{c}_n \tilde{\delta F}_n,
\eeq
where
\begin{align}
\elabel{Sigma}
\Sigma(\omega) &\equiv \int \frac{d^3k}{(2\pi)^3}\frac{1}{Dk^2-i\omega}
	= \frac{1}{2\pi^2}\int_0^\infty dk\frac{k^2}{Dk^2-i\omega},
                 \quad \mbox{and}\\
\elabel{V}
V(x,\omega) &\equiv \int \frac{d^3k}{(2\pi)^3}
	\frac{e^{-i\vec{k}\cdot\vec{x}}}{Dk^2-i\omega}
	= \frac{1}{2\pi^2x}\int_0^\infty dk\frac{k\sin(kx)}{Dk^2-i\omega}
\end{align}
are the ``self-energy'' and ``interaction potential'' between cells
mediated by diffusion, respectively \cite{Bialek2005}.  The
simplifications in \erefs{Sigma}{V} come from writing the volume
element $d^3k = k^2\sin\theta\, dk\, d\theta\, d\phi$ in spherical
coordinates aligned with $\vec{x}$, such that
$\vec{k}\cdot\vec{x} = kx\cos\theta$.

We are interested in the low-frequency limits of $\Sigma(\omega)$ and
$V(x,\omega)$.  $V(x,0) = 1/(4\pi Dx)$ is finite, whereas $\Sigma(0)$
diverges.  The divergence stems from the delta functions in the
dynamical equations, which model the cells as point sources.  As in
\cite{Bialek2005}, we regularize the divergence by introducing a
cutoff $\Lambda \sim \pi/a$ at large $k$ to account for the fact
that cells have finite extent $a$, making
$\Sigma(0) \sim (2\pi^2)^{-1} \int_0^\Lambda dk/D = 1/(2\pi aD)$.
This models cells as spheres of diameter $a$, but the exact shape
of the cell will not be important for the limits we
take.
These expressions allow us to write \eref{rf} as
\beq \elabel{rf2} \sum_{n'=1}^NL_{nn'} \tilde{\delta r}_{n'} =
\frac{\alpha \bar{c}_n}{\mu} \left(1+i\omega\tau_1\right)\tilde{\delta
  F}_n, \eeq where \beqn \elabel{tau1}
\tau_1 &\equiv& \frac{1}{\mu}+\frac{\alpha/\mu}{2\pi aD}, \\
\elabel{L}
L_{nn'} &\equiv& \delta_{nn'} + \frac{z(1-\delta_{nn'})}{|n-n'|}, \\
\elabel{z} z &\equiv& -i\omega\frac{\alpha/\mu}{4\pi aD}.  \eeqn In
writing \eref{rf2}, we have assumed that $\omega\tau_1$ is small. This
assumption is valid for integration times $T = 2\pi/\omega$ much
longer than $\tau_1$. The quantity $\tau_1$ is the receptor
equilibration time: it is the time it takes for a signal molecule to
unbind from the receptors and diffuse away from the cell into the bulk
\cite{Agmon1990}. Its first term $\mu^{-1}$ is the intrinsic receptor
unbinding time, and its second term $K/4\pi\sigma D$ (where
$K = \alpha/\mu$ is the equilibrium constant and $\sigma = a/2$ is the
cell radius) accounts for rebinding events that occur before the
molecule diffuses away from the cell completely
\cite{Kaizu2014}. Either term can dominate: the first term dominates
if the intrinsic association rate $\alpha$ is much smaller than the
diffusion-limited association rate $4\pi\sigma D$, because then the
molecule rarely rebinds. Conversely, the second term dominates if
$\alpha$ is much larger than $4\pi\sigma D$, because then rebinding is
frequent, and comprises most of the escape time.  We see from
\erefs{tau1}{z} that $|z| < \omega\tau_1$; therefore, we also treat
$z$ as a small parameter.

Solving \eref{rf2} for $\tilde{\delta r}_{n}$ requires inverting the matrix
$L_{nn'}$.
This matrix is a Toeplitz matrix (a matrix with constant diagonals),
which has a known inversion algorithm \cite{Zohar1969}. Since
$L_{nn'}$ is also symmetric, it is completely specified by its first
row $[1\ \rho_1\ \rho_2\ \dots\ \rho_{N-1}]$. The inversion is
performed recursively as follows.  First one introduces $N-1$ scalars
$h_1$, $h_2$, $\dots$, $h_{N-1}$ and $N-1$ column vectors
$\vec{q}^{(1)}$, $\vec{q}^{(2)}$, $\dots$, $\vec{q}^{(N-1)}$.  These
are initialized as \beq h_1 = 1-(\rho_1)^2, \qquad \vec{q}^{(1)} =
[-\rho_1], \eeq and updated as \beq h_{k+1} = h_k -
\frac{(\zeta_k)^2}{h_k}, \qquad \vec{q}^{(k+1)} = \left[\begin{matrix}
    \vec{q}^{(k)} - \zeta_k\hat{\vec{q}}^{(k)}/h_k\\
    -\zeta_k/h_k\\
\end{matrix}\right],
\eeq where
$\zeta_k \equiv \rho_{k+1} + \sum_{\ell=1}^k \rho_\ell
\hat{q}^{(k)}_\ell$,
and $\hat{\vec{q}}$ is $\vec{q}$ with the elements in reverse order.
Then the inverse is written in terms of the final quantities
$h_{N-1}\equiv H$ and $\vec{q}^{(N-1)}\equiv\vec{Q}$.  The upper left
element is \beq \elabel{inv1} L^{-1}_{11} = 1/H, \eeq the rest of the
first row and column are \beq \elabel{inv2} L^{-1}_{i+1,1} =
L^{-1}_{1,i+1} = Q_i/H \qquad (1\le i\le N-1), \eeq and the diagonals
are calculated recursively from the first row and column as \beq
\elabel{inv3} L^{-1}_{i+1,j+1} = L^{-1}_{ij} +
(Q_iQ_j-\hat{Q}_i\hat{Q}_j)/H \qquad (1\le \{i,j\}\le N-1).  \eeq

We now apply this algorithm to \eref{L}, keeping only terms up to
first order in the small quantity $z$.  From \eref{L} we have
$\rho_j = z/j$. The initial values are \beq h_1 = 1-z^2 \approx 1,
\qquad \vec{q}^{(1)} = [-z].  \eeq Using
$\zeta_1 = z/2 + (z)(z) \approx z/2$, the first recursive step gives
\beq h_2 = 1 - \frac{(z/2)^2}{1}\approx 1, \qquad \vec{q}^{(2)} =
\left[\begin{matrix}
    -z - (z/2)(-z)/(1)\\
    -(z/2)/(1)\\
\end{matrix}\right]
\approx
\left[\begin{matrix}
	-z\\
	-z/2\\
\end{matrix}\right].
\eeq Continuing the recursion establishes the general formulas \beq
h_k = 1, \qquad \vec{q}^{(k)}_j = -z/j \qquad (j \le k), \eeq from
which we extract the final values, \beq H = 1, \qquad Q_j = -z/j.
\eeq These values immediately provide the first row and column of the
inverse according to \erefs{inv1}{inv2}.  Then noting that the second
term on the right hand side of \eref{inv3} is second order in $z$, that
equation implies that the diagonals of the inverse are constant.  We
therefore have the inverse to first order in $z$,
\beq \elabel{Linv}
L^{-1}_{nn'} = \delta_{nn'} - \frac{z(1-\delta_{nn'})}{|n-n'|}.  \eeq

The inverse allows us to solve \eref{rf2} for $\tilde{\delta r}_n$,
\beq
\tilde{\delta r}_n = \sum_{n'=1}^N G_{nn'} \tilde{\delta F}_{n'}
\eeq
where
\beq
G_{nn'}(\omega) = \frac{\alpha}{\mu}
	\left(1+i\omega\tau_1\right)
	L^{-1}_{nn'}\bar{c}_{n'}
= \begin{cases}
\frac{\alpha\bar{c}_n}{\mu}
	\left[1+i\frac{\omega}{\mu}\left(1+\frac{\alpha}{2\pi aD}\right)\right]
	& n'=n,\\
\frac{\alpha^2\omega\bar{c}_{n'}}{4\pi aD\mu^2|n-n'|}
	\left[i-\frac{\omega}{\mu}\left(1+\frac{\alpha}{2\pi aD}\right)\right]
	& n'\ne n,
\end{cases}
\eeq
is the generalized susceptibility. The fluctuation-dissipation theorem then gives the power spectrum,
\beq
\elabel{S}
S^{rr}_{nn'}(\omega) = \frac{2}{\omega}{\rm Im}[G_{nn'}(\omega)]
= \frac{2\alpha\bar{c}_{n'}}{\mu^2}
\begin{cases}
\left(1+\frac{\alpha}{2\pi aD}\right)
	& n'=n,\\
\frac{\alpha}{4\pi aD}\frac{1}{|n-n'|}
	& n'\ne n,
\end{cases}
\eeq
as in \eref{Sr}.

\section{SNR in the many-cell, strong-communication limit}
\alabel{limit}

The variance of the readout is given by \eref{noise},
\beq
\elabel{noiseA}
(\delta\Delta_N)^2 = \frac{\beta^2}{\nu^2}\left[\frac{S^{rr}_{NN}(0)}{T}
	+ \sum_{n,n'=1}^N K_{N-n} K_{N-n'} \frac{S^{rr}_{nn'}(0)}{T}
	- 2\sum_{n=1}^N K_{N-n} \frac{S^{rr}_{Nn}(0)}{T} \right]
	+ \frac{2}{\nu T}\left( \bar{x}_N + K_0\bar{y}_N \right),
\eeq
where $\bar{x}_N = \alpha\beta\bar{c}_N/\mu\nu$ and
$\bar{y}_N = \alpha\beta\sum_{n=0}^{N-1} K_n\bar{c}_{N-n}/\mu\nu$.  In
the limit of many cells ($N \gg 1$) and strong communication
($\gamma \gg \nu$), the kernel takes the approximate form
\beq
\elabel{KA}
K_n \approx \frac{1}{n_0}e^{-n/n_0},
\eeq
where $n_0 \equiv \sqrt{\gamma/\nu} \gg 1$ is the communication
lengthscale. Using \erefs{S}{KA}, we evaluate \eref{noiseA} term by
term.

The first, fourth, and fifth terms in \eref{noiseA} are straightforward,
\beqn
\elabel{1}
\frac{\beta^2}{\nu^2}\frac{S^{rr}_{NN}(0)}{T}
	&=& \frac{2}{\mu T}\frac{\alpha\beta^2}{\mu\nu^2}\bar{c}_N
	+ \frac{1}{\pi aDT}\frac{\alpha^2\beta^2}{\mu^2\nu^2}\bar{c}_N, \\
\elabel{4}
\frac{2}{\nu T} \bar{x}_N
	&=& \frac{2}{\nu T}\frac{\alpha\beta}{\mu\nu}\bar{c}_N, \\
\elabel{5}
\frac{2}{\nu T} K_0 \bar{y}_N
	&=& \frac{2}{\nu T}\frac{\alpha\beta}{\mu\nu}\frac{\bar{c}_{N-n_0}}{n_0}.
\eeqn
In the last equation we use the limit of large $N$ to approximate the sum in $\bar{y}_N$ as an integral,
\beq
\elabel{int}
\sum_{n=0}^{N-1} K_n\bar{c}_{N-n} \approx \int_0^\infty dn\ \frac{1}{n_0}e^{-n/n_0} (\bar{c}_N - nag)
	= \bar{c}_N - n_0ag = \bar{c}_{N-n_0}.
\eeq

The third term in \eref{noiseA} we split in two, \beq
-2\frac{\beta^2}{\nu^2} \sum_{n=1}^N K_{N-n} \frac{S^{rr}_{Nn}(0)}{T}
= -2\frac{\beta^2}{\nu^2} \left[ K_0 \frac{S^{rr}_{NN}(0)}{T} +
  \sum_{n=1}^{N-1} K_{N-n} \frac{S^{rr}_{Nn}(0)}{T}\right].  \eeq The
first of these is like \eref{1} \beq -2\frac{\beta^2}{\nu^2} K_0
\frac{S^{rr}_{NN}(0)}{T} = -\frac{4}{\mu
  T}\frac{\alpha\beta^2}{\mu\nu^2}\frac{\bar{c}_N}{n_0} - \frac{2}{\pi
  aDT}\frac{\alpha^2\beta^2}{\mu^2\nu^2}\frac{\bar{c}_N}{n_0}.  \eeq
The second evaluates to \beqn -2\frac{\beta^2}{\nu^2} \sum_{n=1}^{N-1}
K_{N-n} \frac{S^{rr}_{Nn}(0)}{T} &=& -2\frac{\beta^2}{\nu^2}
\sum_{n=1}^{N-1} \frac{1}{n_0}e^{-(N-n)/n_0}
\frac{2\alpha}{\mu^2T}[\bar{c}_N-(N-n)ag]\frac{\alpha}{4\pi aD}\frac{1}{N-n} \\
&=& - \frac{1}{\pi aDT}\frac{\alpha^2\beta^2}{\mu^2\nu^2}
\left[\frac{\bar{c}_N}{n_0}\sum_{j=1}^{N-1}\frac{e^{-j/n_0}}{j}
  - \frac{ag}{n_0}\sum_{j=1}^{N-1} e^{-j/n_0}\right] \\
&\approx& - \frac{1}{\pi aDT}\frac{\alpha^2\beta^2}{\mu^2\nu^2}
\left[\frac{\bar{c}_N}{n_0}\log n_0 - ag \right], \eeqn where the last
step again uses the integral approximation for large $N$. In
particular, we have written \beq \elabel{Gamma}
\sum_{j=1}^{N-1}\frac{e^{-j/n_0}}{j} \approx \int_1^\infty dj\
\frac{e^{-j/n_0}}{j} = \Gamma(0,1/n_0) \approx \log n_0 - \gamma_e
\approx \log n_0, \eeq using the small-argument limit of the upper
incomplete Gamma function. The last step assumes that $\log n_0$ is
much larger than the Euler-Mascheroni constant
$\gamma_e \approx 0.577$. Numerically, we have checked that
\eref{Gamma} is valid for $5 \lesssim n_0 \ll N$.
  Altogether, the third term in \eref{noiseA} is then
\beqn
-2\frac{\beta^2}{\nu^2} \sum_{n=1}^N K_{N-n} \frac{S^{rr}_{Nn}(0)}{T}
	&=& -\frac{4}{\mu T}\frac{\alpha\beta^2}{\mu\nu^2}\frac{\bar{c}_N}{n_0}
	- \frac{2}{\pi aDT}\frac{\alpha^2\beta^2}{\mu^2\nu^2}
		\left[\frac{\bar{c}_N}{n_0}+\frac{\bar{c}_N}{2n_0}\log n_0 - \frac{ag}{2}\right] \\
\elabel{3}
	&\approx& -\frac{4}{\mu T}\frac{\alpha\beta^2}{\mu\nu^2}\frac{\bar{c}_N}{n_0}
	- \frac{2}{\pi aDT}\frac{\alpha^2\beta^2}{\mu^2\nu^2}\frac{\log n_0}{2n_0}\bar{c}_N,
\eeqn
where the second step uses $\bar{c}_N-agn_0/2 = \bar{c}_{N-n_0/2} < \bar{c}_N$ and assumes $\log n_0 \gg 2$.

The second term in \eref{noiseA} we also split in two,
\beq
\frac{\beta^2}{\nu^2}\sum_{n,n'=1}^N K_{N-n} K_{N-n'} \frac{S^{rr}_{nn'}(0)}{T}
	= \frac{\beta^2}{\nu^2}\left[ \sum_{n=1}^N K_{N-n}^2 \frac{S^{rr}_{nn}(0)}{T}
	+ \sum_{n=1}^N\sum_{n'\neq n} K_{N-n} K_{N-n'} \frac{S^{rr}_{nn'}(0)}{T} \right].
\eeq
The first of these is straightforward to evaluate with the integral approximation in \eref{int},
\beqn
\frac{\beta^2}{\nu^2}\sum_{n=1}^N K_{N-n}^2 \frac{S^{rr}_{nn}(0)}{T}
	&=& \frac{\beta^2}{\nu^2}\sum_{j=0}^{N-1} \frac{1}{n_0^2}e^{-2j/n_0}
		\frac{2\alpha\bar{c}_{N-j}}{\mu^2T}\left(1+\frac{\alpha}{2\pi aD} \right) \\
	&\approx& \frac{2}{\mu T}\frac{\alpha\beta^2}{\mu\nu^2}\frac{\bar{c}_{N-n_0/2}}{2n_0}
	+ \frac{1}{\pi aDT}\frac{\alpha^2\beta^2}{\mu^2\nu^2}\frac{\bar{c}_{N-n_0/2}}{2n_0}.
\eeqn
The second can be evaluated in two parts,
\beq
\elabel{AB}
\frac{\beta^2}{\nu^2}\sum_{n=1}^N\sum_{n'\neq n} K_{N-n} K_{N-n'} \frac{S^{rr}_{nn'}(0)}{T}
	= \frac{1}{2\pi aDT}\frac{\alpha^2\beta^2}{\mu^2\nu^2}
		\Bigg[\frac{\bar{c}_N}{n_0^2} \underbrace{\sum_{j=0}^{N-1}\sum_{j'\neq j}\frac{e^{-(j+j')/n_0}}{|j-j'|}}_A
		- \frac{ag}{n_0^2} \underbrace{\sum_{j=0}^{N-1}\sum_{j'\neq j}\frac{j'e^{-(j+j')/n_0}}{|j-j'|}}_B \Bigg].
\eeq
Notice that $B = -\partial_uA/2$, where $u \equiv 1/n_0$, so that we only need to evaluate $A$. We split $A$ into two equal components,
\beqn
A &=& \sum_{j=0}^{N-1}\sum_{j'=0}^{j-1}\frac{e^{-(j+j')/n_0}}{j-j'}
		+ \sum_{j=0}^{N-1}\sum_{j'=j+1}^{N-1}\frac{e^{-(j+j')/n_0}}{j'-j} \\
	&=& 2\sum_{j=0}^{N-1}\sum_{j'=0}^{j-1}\frac{e^{-(j+j')/n_0}}{j-j'},
\eeqn
and rewrite it in terms of $k = j + j'$ and $\ell = j-j'$,
\beq
A = 2\sum_{\ell=1}^{N-1}\sum_{k=\ell,\ell+2,\ell+4,\dots}^{2(N-1)-\ell}\frac{e^{-k/n_0}}{\ell}.
\eeq
We approximate with integrals for large $N$, accounting for the fact that $k$ has support on only half of the integers in its range,
\beqn
A &\approx& 2\int_1^N d\ell\ \frac{1}{2} \int_\ell^{2(N-1)-\ell} dk\ \frac{e^{-k/n_0}}{\ell} \\
\elabel{Aint}
&=& n_0\int_1^N d\ell\ \frac{e^{-\ell/n_0}}{\ell} - n_0e^{-2N}\int_1^N d\ell\ \frac{e^{\ell/n_0}}{\ell}.
\eeqn
The first integral is approximately $\log n_0$ by \eref{Gamma}. The second integral evaluates to ${\rm Ei}(N/n_0)-{\rm Ei}(1/n_0)$, where Ei is the exponential integral function, whose large- and small-argument limits are ${\rm Ei}(N/n_0) \approx e^{N/n_0}/(N/n_0)$ and ${\rm Ei}(1/n_0) \approx -\log n_0$, respectively. Thus, the second term in \eref{Aint} vanishes exponentially with $N$, and we have
\beqn
A &=& n_0\log n_0, \\
B &=& \frac{n_0^2}{2}(1+\log n_0) \approx \frac{n_0^2}{2}\log n_0,
\eeqn
making the term in brackets in \eref{AB} equal to $(\bar{c}_N - agn_0/2)(\log n_0)/n_0 = (\bar{c}_{N-n_0/2}\log n_0)/n_0$.
Altogether, the second term in \eref{noiseA} is then
\beqn
\frac{\beta^2}{\nu^2}\sum_{n,n'=1}^N K_{N-n} K_{N-n'} \frac{S^{rr}_{nn'}(0)}{T}
	&=& \frac{2}{\mu T}\frac{\alpha\beta^2}{\mu\nu^2}\frac{\bar{c}_{N-n_0/2}}{2n_0}
	+ \frac{1}{\pi aDT}\frac{\alpha^2\beta^2}{\mu^2\nu^2}
	\left[ \frac{\bar{c}_{N-n_0/2}}{2n_0} + \frac{\log n_0}{2n_0}\bar{c}_{N-n_0/2} \right] \\
\elabel{2}
	&\approx& \frac{2}{\mu T}\frac{\alpha\beta^2}{\mu\nu^2}\frac{\bar{c}_{N-n_0/2}}{2n_0}
	+ \frac{1}{\pi aDT}\frac{\alpha^2\beta^2}{\mu^2\nu^2}
	\frac{\log n_0}{2n_0}\bar{c}_{N-n_0/2},
\eeqn
where the second step assumes $\log n_0 \gg 1$.

Finally, collecting the terms in \erefn{1}{5}, \ref{eq:3}, and \ref{eq:2}, the variance in \eref{noiseA} becomes
\begin{align}
\elabel{noise_full}
(\delta\Delta_N)^2 = \left(\frac{\alpha\beta}{a^3\mu\nu}\right)^2
	\Bigg[ &\underbrace{\frac{a^2}{\pi DT} \left(\bar{c}_N + \frac{\log n_0}{2n_0}\bar{c}_{N-n_0/2}
		- 2\frac{\log n_0}{2n_0}\bar{c}_N\right)a^3}_{{\rm {\bf Extrinsic}\ noise\ from\ } c,\
			{\rm propagated\ to\ } x\ {\rm and\ } y} \nonumber\\
	& + \underbrace{\frac{2}{\mu T}\frac{a^3\mu}{\alpha} \left( \bar{c}_N + \frac{\bar{c}_{N-n_0/2}}{2n_0}
		- 2\frac{\bar{c}_N}{n_0} \right)a^3}_{{\rm {\bf Intrinsic}\ noise\ in\ } r,\
			{\rm propagated\ to\ } x\ {\rm and\ } y} \nonumber\\
	& + \underbrace{\frac{2}{\nu T}\frac{a^3\mu}{\alpha}\frac{\nu}{\beta}
		\left(\bar{c}_N\right)a^3}_{{\rm {\bf Intrinsic}\ noise\ in\ } x}
	+ \underbrace{\frac{2}{\nu T}\frac{a^3\mu}{\alpha}\frac{\nu}{\beta}
		\left(\frac{\bar{c}_{N-n_0}}{n_0}\right)a^3}_{{\rm {\bf Intrinsic}\ noise\ in\ } y} \Bigg].
\end{align}
The last line contains the intrinsic noise from $x_N$ and $y_N$. For example, the first term in this line is $(\delta x_N)^2 = (2/\nu T)(\alpha\beta\bar{c}_N/\mu\nu) = 2\bar{x}_N/\nu T$; the relative noise $(\delta x_N/\bar{x}_N)^2 = 2/\nu T\bar{x}_N$ then decreases with the number of molecules $\bar{x}_N\times \nu T$ that are turned over in time $T$, as expected for intrinsic counting noise. The second term in this line is for $y$ and is similar, except that since the global species is exchanged over roughly $n_0$ cells, more molecules are counted and the noise is reduced by a factor $n_0$. The second line contains the intrinsic noise in $r$, propagated to $x$ and $y$. The third line contains the noise in $c$, propagated to $x$ and $y$, which we deem extrinsic, since it originates in the environment and is not under direct control of the cells.

Importantly, the intrinsic noise terms in \eref{noise_full} are reducible by increasing the numbers of receptors and local and global species molecules. These molecule numbers are set by the gain factors $\bar{r}_N/a^3\bar{c}_N = \alpha/a^3\mu$ and $\bar{x}_N/\bar{r}_N = \beta/\nu$, and indeed, we see that the second and third lines in \eref{noise_full} vanish as the gain factors grow large. In this limit we are left with only the lower bound set by the extrinsic noise,
\beq
(\delta\Delta_N)^2 \ge \left(\frac{\alpha\beta}{\mu\nu}\right)^2
	\frac{1}{\pi aDT} \left(\bar{c}_N + \frac{\log n_0}{2n_0}\bar{c}_{N-n_0/2}
		- 2\frac{\log n_0}{2n_0}\bar{c}_N\right).
\eeq
Dividing by the square of the mean $\bar{\Delta}_N = \alpha\beta n_0ag/\mu\nu$ produces \erefs{limit}{ceff}.

\section{Exact SNR for regional excitation--global inhibition (REGI)}
\alabel{regi}

In the REGI strategy, both messengers X and Y are exchanged between
cells, at rates $\gamma_x$ and $\gamma_y$, respectively. This results
in a straightforward generalization of the expression for the
signal-to-noise ratio (SNR). Specifically, the mean becomes (compare
to \erefs{signal}{K} in the main text) \beq \bar{\Delta}_N =
\frac{\alpha\beta}{\mu\nu}\left( \sum_{n=0}^{N-1} K^x_n \bar{c}_{N-n}
  - \sum_{n=0}^{N-1} K^y_n \bar{c}_{N-n} \right), \eeq where now there
are two communication kernels, \beqn K^x_n =
\frac{\sum_{j=0}^{N-n-1}{N-n-1+j \choose 2j}(\nu/\gamma_x)^j}
{\sum_{\ell=0}^{N-1}{N+\ell \choose 2\ell+1}(\nu/\gamma_x)^\ell}, \\
K^y_n = \frac{\sum_{j=0}^{N-n-1}{N-n-1+j \choose 2j}(\nu/\gamma_y)^j}
{\sum_{\ell=0}^{N-1}{N+\ell \choose 2\ell+1}(\nu/\gamma_y)^\ell}.
\eeqn The variance becomes (compare to \eref{noise}) \beq
(\delta\Delta_N)^2 = \frac{\beta^2}{\nu^2} \left[ \sum_{nn'}
  \left(K^x_{N-n} K^x_{N-n'} + K^y_{N-n} K^y_{N-n'} - 2K^x_{N-n}
    K^y_{N-n'} \right) \frac{S^{rr}_{nn'}(0)}{T} \right] +
\frac{2}{\nu T}\left( K^x_0\bar{x}_N + K^y_0\bar{y}_N \right), \eeq
where the bound receptor power spectrum $S^{rr}_{nn'}(\omega)$ remains
the same as in \eref{Sr} (or equivalently \eref{S}). The SNR is then $(\bar{\Delta}_N/\delta\Delta_N)^2$.

The SNR has a maximum as a function of the rate ratio
$\gamma_x/\gamma_y$. The location of the maximum $\gamma_x^*/\gamma_y$
must lie between $0$ and $1$. At $\gamma_x/\gamma_y = 0$, the X
messenger is not exchanged, and we recover the SNR of the local
excitation--global inhibition (LEGI) strategy, which is a limiting
case of REGI. At $\gamma_x/\gamma_y = 1$, there is no difference
between X and Y, and the signal (and therefore the SNR) is $0$. The
exact location of $\gamma_x^*/\gamma_y$ depends on factors that are
specific to the system, e.g.\ the concentration profile $\bar{c}_n$,
the environmental and system parameters, and the measurement location
($n = 1$ vs.\ $n=N$).

We illustrate the dependence of $\gamma_x^*/\gamma_y$ on particular
system parameters, namely the gain factors $\alpha/a^3\mu$ and
$\beta/\nu$. For this, \fref{gain} shows the dependence on SNR on
$\gamma_x/\gamma_y$ as we vary the gain factors $\alpha/a^3\mu$ (A)
and $\beta/\nu$ (B). In both cases, we see that the optimal rate ratio
$\gamma_x^*/\gamma_y$ increases with increasing gain. This is because
increasing either gain factor increases the number of internal
messenger molecules. With more molecules, the system can afford to
increase $\gamma_x$ while maintaining the same difference in molecule
number $\Delta_N$ in the $N$th cell. The increase in $\gamma_x$
enhances the spatial averaging by the X messenger, and thus reduces
the noise in the estimate of $\bar{c}_N$. Therefore, we see in
\fref{gain} that the maximal SNR occurs at a higher $\gamma_x^*$ value
as either gain is increased.

\begin{figure}
	\begin{center}
		\includegraphics[width=0.95\textwidth]{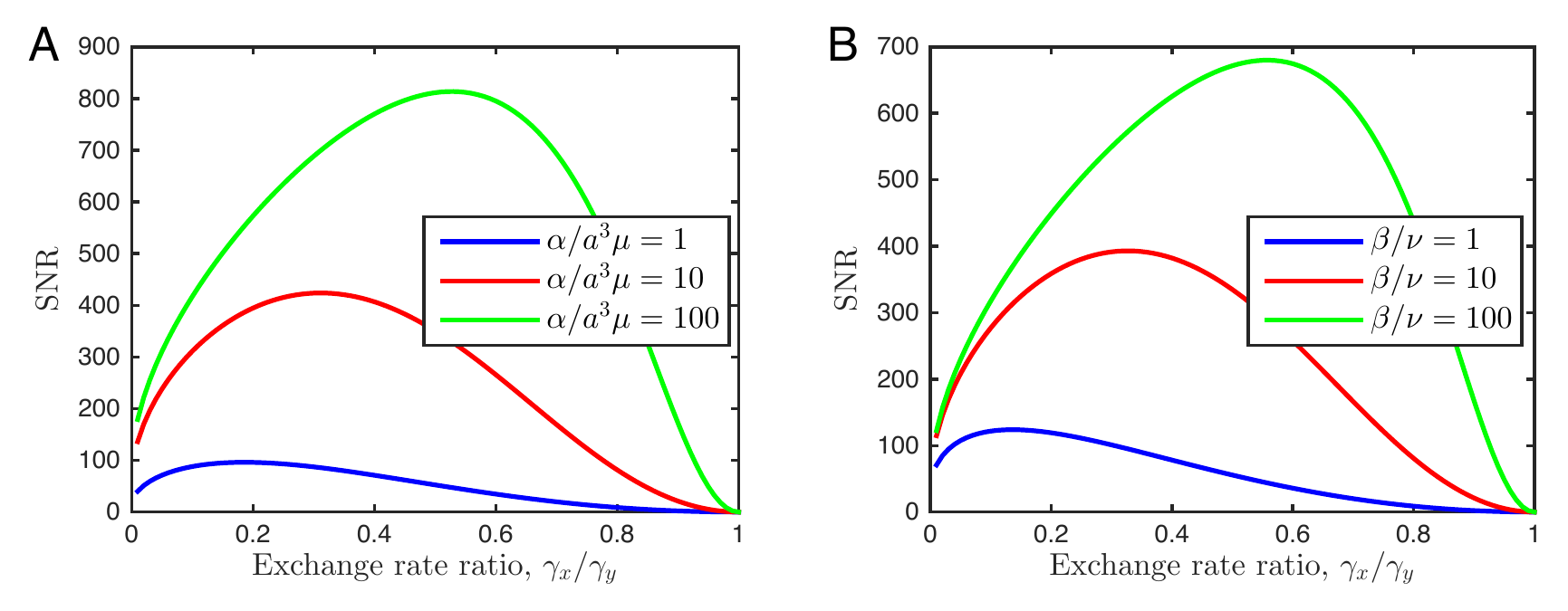}
	\end{center}
	\caption{ \flabel{gain}{\bf Dependence of the optimal exchange
            rate ratio on system parameters in the regional
            excitation--global inhibition (REGI) strategy.} The
          signal-to-noise ratio (SNR) has a maximum at a particular
          rate ratio $\gamma_x^*/\gamma_y$, which increases as a
          function of either gain factor, (A) $\alpha/a^3\mu$, or (B)
          $\beta/\nu$. Parameters are $a = 10\ \mu$m, $\bar{c}_N = 1$
          nM, $g = 1$ nM/mm, $D = 50\ \mu$m$^2$/s, $T = 10$ s,
          $\mu = \nu = 1$ s$^{-1}$, $N = 100$, and
          $n_y = \sqrt{\gamma/\nu} = 10$. In A, $\beta/\nu = 5$, and
          $\alpha/a^3\mu$ is varied as indicated. In B,
          $\alpha/a^3\mu = 5$, and $\beta/\nu$ is varied as indicated.
        }
\end{figure}


\end{document}